\documentclass[aps, pra, twocolumn, showpacs]{revtex4}

\usepackage{amssymb}
\usepackage{graphicx}
\usepackage{dcolumn}
\usepackage{bm}
\usepackage{amsmath, amsfonts}

\begin{document}

\title{Errors in trapped-ion quantum gates due to spontaneous photon scattering}
\author{R.~Ozeri}
\author{W.~M.~Itano}
\author{R.~B.~Blakestad}
\author{J.~Britton}
\author{J.~Chiaverini}
\altaffiliation[Present address: ]{Los Alamos National Laboratory,
Los Alamos, NM 87545, USA.}
\author{J.~D.~Jost}
\author{C.~Langer}
\author{D.~Leibfried}
\author{R.~Reichle}
\altaffiliation[Present address: ]{University of Ulm, 89069 Ulm,
Germany.}
\author{S.~Seidelin}
\author{J.~H.~Wesenberg}
\author{D.~J.~Wineland}
\affiliation{NIST Boulder, Time and Frequency division, Boulder,
Colorado 80305}


\begin{abstract}
We analyze the error in trapped-ion, hyperfine qubit, quantum
gates due to spontaneous scattering of photons from the gate laser
beams. We investigate single-qubit rotations that are based on
stimulated Raman transitions and two-qubit entangling phase-gates
that are based on spin-dependent optical dipole forces. This error
is compared between different ion species currently being
investigated as possible quantum information carriers. For both
gate types we show that with attainable laser powers the
scattering error can be reduced to below current estimates of the
fault-tolerance error threshold.
\end{abstract}

\maketitle

\section{INTRODUCTION}\label{section1}
Quantum bits, or qubits, that are encoded into internal states of
trapped ions are an interesting system for Quantum Information
Processing (QIP) studies \cite{CiracZoller, bible}. Internal
states of trapped ions can be well isolated from the environment,
and very long coherence times are possible \cite{bible,
Wunderlich, HighField, Oxford_LongCoherence}. The internal states
of several ion-qubits can be deterministically entangled, and
quantum gates can be carried out between two ion-qubits
\cite{entanglement, Innsbruck_entanglement, SackettGate,
Didi_gate,Innsbruck_gate,Michigan_gate,Oxford_gate}.

Among different choices of internal states, qubits that are
encoded into pairs of ground state hyperfine or Zeeman states
benefit from negligible spontaneous decay rates \cite{bible}. The
small energy separation between the two states of such qubits
(typically in the radio-frequency or microwave domain) allows for
phase coherence between a local oscillator and a qubit
superposition state over relatively long times
\cite{bible,HighField,Bollinger_longCoherence, Fisk}.

The quantum gates that are performed on hyperfine ion-qubits
typically use laser beams. Since light couples only very weakly to
the electron spin, spin manipulations rely instead on the
spin-orbit coupling of levels that are typically excited non
resonantly through allowed electric dipole transitions. Spin
manipulations therefore require a finite amplitude in the excited
electronic state, and spontaneous scattering of photons from the
laser beams during the gate is inevitable.

Fault-tolerant quantum computation demands that the error in a
single gate is below a certain threshold. Current estimates of the
fault-tolerance error threshold range between $10^{-2}$ to
$10^{-4}$ \cite{Steane_FT, Reichardt_FT, Knill_FT}. These
estimates rely on specific noise models and error-correction
protocols and should be considered as guidelines only. However,
the general view is that for fault tolerance to be practical, the
error probability in quantum gates should be at least as small as
$10^{-4}$. It is, therefore, worth exploring the limitations to
the fidelity of quantum gates performed on trapped ions with laser
light, using this level of error as a guideline
\cite{PlenioKnight}.

In his 1975 paper \cite{Mollow}, Mollow showed that the effect of
a quantum coherent field on an atom is equivalent to that of a
classical field plus a quantum vacuum field. The error due to the
interaction with light can be categorized into two parts. The
first is the error due to noise in classical laser parameters,
such as intensity or phase \cite{bible, Milburn98}. The second
part originates from the quantum nature of the electromagnetic
field and is due to vacuum fluctuations, i.e. the spontaneous
scattering of photons \cite{Wayne, SilberfarbDeutsch}.

\subsection{Ion-qubit levels and transitions}
\begin{figure}[tbp]
\begin{center}
\includegraphics[width=8.6 cm]{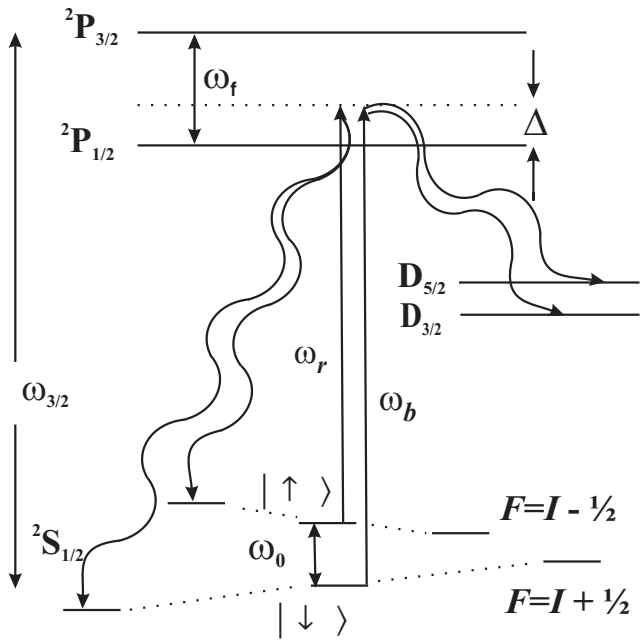}
\end{center}
\caption{Relevant energy levels (not to scale) in an ion-qubit, with
nuclear spin $I$. The $\text{P}_{1/2}$ and $\text{P}_{3/2}$ excited
levels are separated by an angular frequency $\omega_\text{f}$. The
$\text{S}_{1/2}$ electronic ground state consists of two hyperfine
levels $F=I-1/2$ and $F=I+1/2$. The relative energies of these two
levels depends on the sign of the hyperfine constant $A_{\text{hf}}$
and can vary between ion species (in this figure, $A_{\text{hf}}$ is
negative). The qubit is encoded in the pair of $m_F=0$ states of the
two $F$ manifolds separated by an angular frequency $\omega_0$.
Coherent manipulations of the qubit levels are performed with a pair
of laser beams that are detuned by $\Delta$ from the transition to
the $\text{P}_{1/2}$ level, represented by the two straight arrows.
The angular frequency difference between the two beams equals the
angular frequency separation between the qubit levels
$\omega_\text{b}-\omega_\text{r} = \omega_0$. Some ion species have
D levels with energies below the P manifold. Wavy arrows illustrate
examples of Raman scattering events.} \label{Levels}
\end{figure}

Most ion species considered for QIP studies have a single valence
electron, with a $^2\text{S}_{1/2}$ electronic ground state, and
$^2\text{P}_{1/2}$ and $^2\text{P}_{3/2}$ electronic excited
states. Some of the ions also have D levels with lower energy than
those of the excited state P levels. Ions with a non zero nuclear
spin also have hyperfine structure in all of these levels. A small
magnetic field is typically applied to remove the degeneracy
between different Zeeman levels. Here we consider qubits that are
encoded into a pair of hyperfine levels of the $^2\text{S}_{1/2}$
manifold. Figure \ref{Levels} illustrates a typical energy level
structure.

To allow for a straightforward comparison between different ion
species, we investigate qubits that are based on clock
transitions, i.e. a transition between the $|F=I-1/2, m_F =
0\rangle \equiv |\uparrow \rangle$ and $|F=I+1/2, m_F = 0\rangle
\equiv |\downarrow \rangle$ hyperfine levels in the
$\text{S}_{1/2}$ manifold of ions with a half odd-integer nuclear
spin $I$, at a small magnetic field. For this transition, the
total Raman and Rayleigh photon scattering rates, as well as the
Rabi frequency, are independent of $I$, and the comparison between
different ion species depends only on other atomic constants.
Superpositions encoded into these states are also more resilient
against magnetic field noise \cite{Wunderlich,
Oxford_LongCoherence, Michigan_gate}. Even though our quantitative
results apply only to this configuration, for other choices of
hyperfine or Zeeman qubit states (including those with $I=0$) the
results will not change significantly.

Gates are assumed to be driven by pairs of Raman beams detuned by
$\Delta$ from the transition between the $\text{S}_{1/2}$ and the
$\text{P}_{1/2}$ levels (See Fig. \ref{Levels}). We further assume
that the Raman beams are linearly polarized and Raman transitions
are driven by both $\sigma_{+}$ photon pairs and $\sigma_{-}$
photon pairs. The two beams in a Raman pair are designated as red
Raman ($r$) and blue Raman ($b$) by their respective frequencies.
In the following we also assume that $\Delta$ is much larger than
the hyperfine and Zeeman splitting between levels in the ground
and excited states.

The Rabi frequency between the two clock states is \cite{London}
\begin{equation}
\Omega_\text{R}=\frac{g_{b}g_{r}}{3}(b_{-}r_{-}-b_{+}r_{+})\frac{\omega_\text{f}}{\Delta(\Delta-\omega_\text{f})},
\label{Rabi}
\end{equation}
where $g_{b/r}=E_{b/r}|\langle \text{P}_{3/2},F=I+3/2,
m_F=\text{F}|\hat{\textbf{d}}\cdot\hat{\sigma}_{+}\
|\text{S}_{1/2},F=I+1/2, m_F=F \rangle|/2\hbar$, $E_{b/r}$ is the
peak electric field of the $b$ or $r$ beam at the position of the
ion, respectively, and $\hat{\textbf{d}}\cdot\hat{\sigma}_{+}$ is
the right circular component of the electric dipole operator. The
right and left circular polarization components of the $b$ and $r$
beams are $b_{+/-}$ and $r_{+/-}$, respectively. The
$\text{P}_{1/2}$ and $\text{P}_{3/2}$ excited levels are separated
by an angular frequency $\omega_\text{f}$.

\begin{figure}[tbp]
\begin{center}
\includegraphics[width=8.6cm]{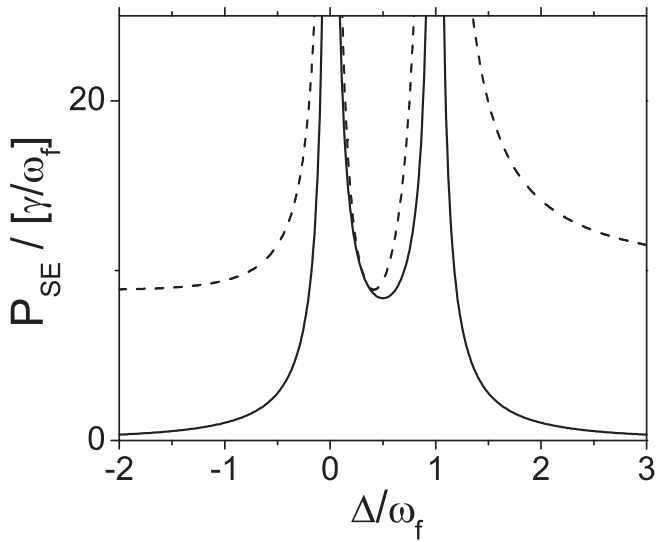}
\end{center}
\caption{The solid line is the probability ($P_{\text{SE}} =
P_{\text{Raman}}$)to scatter a Raman photon (in units of
$\gamma/\omega_\text{f}$) during a $\pi$ rotation vs. the laser
detuning in units of the excited state fine-structure splitting.
The dashed line is the probability for any type of scattering
event  ($P_{\text{SE}} = P_{\text{total}}$) during the pulse vs.
detuning. The Raman scattering probability decays quadratically
with $\Delta$ for $|\Delta| \gg \omega_\text{f}$.}
\label{P_Raman_vs_Detuning}
\end{figure}

The total spontaneous photon scattering rate from these beams is
given by \cite{London}
\begin{equation}
\Gamma_{\text{total}}=\frac{\gamma}{3}\left[
g_{b}^2(b_{-}^2+b_{+}^2)+g_{r}^2(r_{-}^2+r_{+}^2)\right]\left[
\frac{1}{\Delta^2}+ \frac{2}{(\Delta-\omega_\text{f})^2}\right].
\label{total_rate}
\end{equation}
Here $\gamma$ is the natural linewidth of the $\text{P}_{1/2}$ and
$\text{P}_{3/2}$ levels \cite{gRemark}. We now assume linearly
polarized Raman beams with $b_{-}=b_{+}=r_{-}=-r_{+}=1/\sqrt{2}$
\cite{London}. We further assume $g_{b} = g_{r} \equiv g$. The
time for a $\pi$ rotation is
\begin{equation}
\tau_{\pi}=\frac{\pi}{2\Omega_\text{R}} \label{pi_time}
\end{equation}
Combining Eq. (\ref{Rabi}),(\ref{total_rate}) and (\ref{pi_time}),
the probability to scatter a photon during $\tau_{\pi}$ is given
by
\begin{equation}
P_{\text{total}}=(\frac{\pi\gamma}{\omega_\text{f}})\frac{2\Delta^{2}+(\Delta-\omega_\text{f})^2}{|\Delta(\Delta-\omega_\text{f})|}.
\label{photon per pulse}
\end{equation}
The dashed line in Fig. \ref{P_Raman_vs_Detuning} shows
$P_{\text{total}}$ vs. $\Delta$, where the laser detuning is
expressed in units of the excited state fine-structure splitting,
and the scattering probability is given in units of
$\gamma/\omega_\text{f}$. The total scattering probability has a
global minimum of
\begin{equation}
P_{min}=2\sqrt{2}\pi\gamma/\omega_\text{f}, \label{total_min}
\end{equation}
when the laser detuning is between the two fine-structure
manifolds ($\Delta = (\sqrt{2} - 1) \omega_\text{f}$). The
asymptotic value of $P_\text{total}$ for large positive or
negative detuning
\begin{equation}
P_{\infty}=3\pi\gamma/\omega_\text{f}, \label{total_limit}
\end{equation}
is only slightly larger than the global minimum.

Previous studies estimated decoherence by assuming that any photon
scattering will immediately decohere a hyperfine superposition
\cite{London}. Under this assumption the lowest possible gate
error equals $P_{min}$ and ions with a small
$\gamma/\omega_\text{f}$ ratio benefit from a lower gate error
minimum.

Two kinds of off-resonance photon scattering occur in the presence
of multiple ground states. Inelastic Raman scattering which
transfers population between ground states, and Rayleigh elastic
scattering which does not change ground state populations. Since
no energy or angular momentum are exchanged between the photon and
the ion internal degrees of freedom, no information about the
qubit state is carried away by a Rayleigh scattered photon.
Rayleigh scattering, therefore, does not necessarily lead to
decoherence \cite{London, cline94, spont_paper}. This was
experimentally shown in \cite{spont_paper} where, when of equal
rate from both qubit levels, off-resonance Rayleigh scattering of
photons did not affect the coherence of a hyperfine superposition.
Decoherence in the presence of light was shown to be dominated by
Raman scattering. The guideline used in \cite{London} is therefore
overly pessimistic.

In this paper we re-examine the errors due to spontaneous photon
scattering on single-qubit gates (rotations of the equivalent spin
$1/2$ vector on the Bloch sphere) and two-qubit (entanglement)
gates, where the qubits are based on ground state hyperfine levels
and manipulated with stimulated Raman transitions. We compare
between different ion species and examine different Raman laser
parameters. In section \ref{section2} we analyze the contribution
to the gate error due to spontaneous Raman scattering. Following a
Raman scattering event the ion-qubit is projected into one of its
ground states and spin coherence is lost. This error was also
addressed in \cite{Gea-Banacloche}. In section \ref{section3} we
examine the error due to Rayleigh scattering. Rayleigh scattering
error results primarily from the photon's recoil momentum kick. We
show that both types of gate errors can be reduced to small
values, while keeping the gate speed constant, with the use of
higher laser intensity.

\section{RAMAN SCATTERING ERROR}\label{section2}
In a Raman photon scattering event energy and angular momentum are
exchanged between the scattered photon and the ion's internal
degrees of freedom. The polarization and frequency of the
scattered photon (with respect to those of the laser) become
entangled with the ions' internal state. Therefore, after tracing
out the photon degrees of freedom, the ions' spin coherence is
lost. In other words, Raman scattering serves as a measurement of
the ion-qubit internal state. Following a spontaneous Raman
scattering event the ion-qubit is projected into one of the ground
states in the $\text{S}_{1/2}$ manifold. For ions with low lying D
levels, Raman scattering events can also transfer the ion from the
qubit levels into one of the D levels.

\subsection{Single-qubit gate} For a single qubit gate we choose to look at the
fidelity of a $\pi$ rotation around the $x$-axis of the Bloch
sphere, represented by the Pauli operator $\hat{\sigma}_{x}$. This
gate is assumed to be driven by a co-propagating Raman beam pair,
where the frequency difference between the beams equals the
frequency separation between the two qubit states $\omega_0$. It
is straightforward to generalize this case to other rotations.

As a measure of the error in the rotation, we use the fidelity of
the final state (characterized by density matrix
$\hat{\rho}_{\text{final}}$) produced by the erroneous gate as
defined by
\begin{equation}
F = \langle \Psi |\hat{\rho}_{\text{final}}|\Psi\rangle,
\label{fidelity}
\end{equation}
where $|\Psi\rangle$ is the ideal final state. Given an initial
state $|\Psi_{init}\rangle$, $|\Psi\rangle$ can be written as
\begin{equation}
|\Psi\rangle = \hat{\sigma}_{x}|\Psi_{\text{init}}\rangle.
\label{desired}
\end{equation}

In the presence of off-resonance Raman scattering, the density
matrix that describes the state of the qubit after the gate has
the form
\begin{eqnarray}
&& \hat{\rho}_{\text{final}} = (1 -
P_{\text{Raman}})|\Psi\rangle\langle\Psi|+\hat\rho_{\epsilon}.
\label{RhoFinal}
\end{eqnarray}
The erroneous part of the density matrix
$\hat\rho_{\epsilon}=\sum_i w_i |i\rangle\langle i|$, is composed
of projectors into different levels $|i\rangle$ of lower energy.
Here $P_\text{Raman}=\sum_i w_i$ is the probability for a
spontaneous Raman scattering event to occur during the gate.

Note that some Raman scattering events keep the ion within the
qubit manifold. Using Eq. (\ref{fidelity}) the contribution of
$\hat\rho_{\epsilon}$ to the fidelity is positive and not strictly
zero. For simplicity, we neglect this contribution and put a lower
bound on the gate fidelity
\begin{equation}
F \geq 1 - P_{\text{Raman}}. \label{Thefidelity}
\end{equation}
In what follows we assume this expression to be an equality. The
error in the gate due to spontaneous Raman photon scattering is
hence given by the Raman scattering probability
\begin{equation}
\epsilon\equiv1-F = P_{\text{Raman}}.
\end{equation}

We first examine the error due to Raman scattering back into the
$^2\text{S}_{1/2}$ manifold $\epsilon_\text{S}$. The Raman
scattering rate back into the $\text{S}_{1/2}$ manifold is
calculated to be \cite{spont_paper}
\begin{equation}
\Gamma_{\text{Raman}}=\frac{2\gamma}{9}\left[g_{b}^2(b_{-}^2+b_{+}^2)+g_{r}^2(r_{-}^2+r_{+}^2)\right]\left[\frac{\omega_\text{f}}{\Delta(\Delta-\omega_\text{f})}\right]^2.
\label{Raman_rate}
\end{equation}
For the same laser parameters as above, the probability to scatter
a Raman photon during the gate is
\begin{equation}
P_{\text{Raman}}=\frac{2\pi\gamma}{3}\frac{\omega_\text{f}}{|\Delta(\Delta-\omega_\text{f})|}=\epsilon_\text{S}.
\label{Raman_prob1}
\end{equation}
The solid line in Fig. \ref{P_Raman_vs_Detuning} shows $P_{Raman}$
vs. $\Delta$. The Raman scattering probability decays
quadratically with $\Delta$ for $|\Delta| \gg \omega_\text{f}$.
Qualitatively, this is because Raman scattering involves a
rotation of the electron spin. Electron spin rotations are
achieved through the spin-orbit coupling in the excited state.
This coupling has opposite sign contributions from the two
fine-structure levels. Therefore as we detune far compared to the
fine structure splitting, those two contributions nearly cancel
\cite{cline94}.

Using Eq. (\ref{Rabi}) we can write
\begin{equation}
P_{\text{Raman}}=\frac{2\pi\gamma|\Omega_\text{R}|}{g^2}.
\label{Raman_prob2}
\end{equation}
The ratio $\gamma/g^2$ can be expressed in terms of atomic
constants, and the peak electric field amplitude $E$ at the
position of the ion \cite{gRemark}
\begin{equation}
\frac{\gamma}{g^2}=\frac{4\hbar\omega_{3/2}^3}{3\pi\epsilon_{0}c^3E^2}.
\label{gamma_g2_ratio}
\end{equation}
Here $\omega_{3/2}$ is the frequency of the transition between the
$\text{S}_{1/2}$ and the $\text{P}_{3/2}$ levels (D2 line),
$\epsilon_{0}$ is the vacuum permittivity, and $c$ is the speed of
light. Assuming Gaussian laser beams, at the center of the beam
\begin{equation}
E^2=\frac{4\mathcal{P}}{\pi w_{0}^2 c\epsilon_{0}}.
\label{Amplitude2}
\end{equation}
Here $\mathcal{P}$ is the power in each of the Raman beams and
$w_{0}$ is the beam waist at the position of the ion. The
probability to scatter a single Raman photon can be written as
\begin{equation}
P_{\text{Raman}}=\frac{2\pi|\Omega_{\text{R}}|\hbar\omega_{3/2}^3
w_{0}^2}{3c^2\mathcal{P}}=\epsilon_\text{S}. \label{Raman_prob3}
\end{equation}

\begin{table*}[tbp]
\begin{center}
  \begin{tabular}{|l|c|c|c|c|c|c|c|}
  \hline
  Ion&I&$\gamma/2\pi$ (MHz)&$\omega_0/2\pi$ (GHz)&$\omega_\text{f}/2\pi$ (THz)& $\lambda_{1/2}$(nm)&$\lambda_{3/2}$(nm)&$f^{-1}$\\
  \hline
  \hline
  $^{9}$Be$^+$& 3/2 & 19.6 & 1.25 & 0.198 & 313.1 & 313.0 & N.A. \\
  \hline
  $^{25}$Mg$^+$& 5/2 & 41.3 & 1.79 & 2.75 & 280.3 & 279.6 & N.A. \\
  \hline
  $^{43}$Ca$^+$& 7/2 & 22.5 & 3.23 & 6.68 & 396.8 & 393.4 & 17 \\
  \hline
  $^{67}$Zn$^+$& 5/2 & 62.2 & 7.2 & 27.8 & 206.2 & 202.5 & N.A.\\
  \hline
  $^{87}$Sr$^+$& 9/2 & 21.5 & 5.00 & 24.0 & 421.6 & 407.8 & 14 \\
  \hline
  $^{111}$Cd$^+$& 1/2 & 50.5 & 14.53 & 74.4 & 226.5 & 214.4 & N.A. \\
  \hline
  $^{137}$Ba$^+$& 3/2 & 20.1 & 8.04 & 50.7 & 493.4 & 455.4 & 3\\
  \hline
  $^{171}$Yb$^+$& 1/2 & 19.7 & 12.64 & 99.8 & 369.4 & 328.9 & 290\\
  \hline
  $^{199}$Hg$^+$& 1/2 & 54.7 & 40.51 & 273.4 & 194.2 & 165.0 & 700\\
  \hline

  \end{tabular}
\caption{A list of atomic constants of several of the ions
considered for quantum information processing. Here I is the
nuclear spin, $\gamma$ is the natural linewidth of the
$\text{P}_{1/2}$ level \cite{Be_lifetime, Mg_lifetime,
Ca_lifetime, Zn_lifetime, Sr_Ba_lifetime, Cd_lifetime,
Yb_lifetime, Hg_lifetime}, $\omega_0$ is the frequency separation
between the two qubit states set by the hyperfine splitting of the
$\text{S}_{1/2}$ level \cite{Be_hyperfine, Mg_hyperfine,
Ca_hyperfine, Zn_hyperfine, Sr_hyperfine, Cd_hyperfine,
Ba_hyperfine, Yb_hyperfine, Hg_hyperfine}, $\omega_\text{f}$ is
the fine-structure splitting \cite{NIST Data base},
$\lambda_{1/2}$ and $\lambda_{3/2}$ are the wavelengths of the
transitions between the $\text{S}_{1/2}$ and the $\text{P}_{1/2}$
and $\text{P}_{3/2}$ levels \cite{NIST Data base}, respectively.
The branching ratio of decay from the P levels to the D and the S
levels is $f$ \cite{D1 remark,D2 remark, f1, f2, f3, f4, f5}.}
\label{Table0}
\end{center}
\end{table*}

This result is essentially the same as Eq. (5) of
\cite{Gea-Banacloche}. We can now rearrange this expression to put
an upper bound on the required power for a desired gate speed and
error
\begin{equation}
\mathcal{P}=\frac{2\pi}{3\epsilon_\text{S}}(\frac{2\pi
w_{0}}{\lambda_{3/2}})^2\hbar\omega_{3/2}|\Omega_\text{R}|,
\label{Power}
\end{equation}
where $\lambda_{3/2}=c/\omega_{3/2}$.

Assume that the ratio of the beam waist to the transition
wavelength is constant for different ion species. In this case,
the power needed to obtain a given Rabi frequency and to keep the
error below a given value would scale linearly with the optical
transition frequency. A more realistic assumption might be that
the Raman beam waist is not diffraction limited and is determined
by other experimental considerations, such as the inter-ion
distance in the trap or beam pointing fluctuations. In this case,
assuming that $w_{0}$ is constant, the required power would scale
as the optical transition frequency cubed. Either way, ion species
with optical transitions of longer wavelength are better suited in
the sense that less power is required for the same gate speed and
error requirements. In addition, high laser power is typically
more readily available at longer wavelengths. Finally, we note
that the error is independent of the fine-structure splitting as
long as we have sufficient power to drive the transition. The
transition wavelengths of different ions are listed in Table
\ref{Table0}.

\begin{figure}[tbp]
\begin{center}
\includegraphics[width=8.6cm]{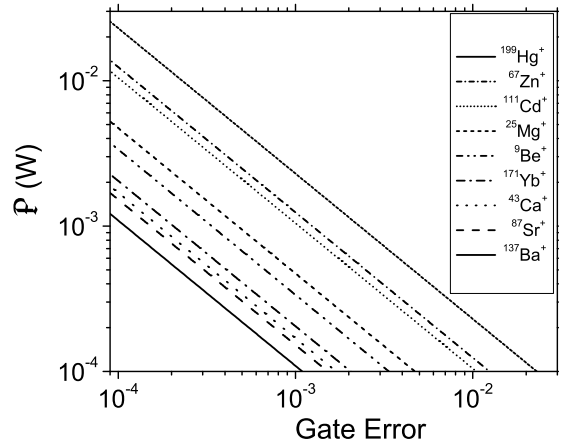}
\end{center}
\caption{Laser power in each of the Raman beams vs. the error in a
single ion gate ($\pi$ rotation) due to Raman scattering back into
the $\text{S}_{1/2}$ manifold (obtained using Eq. (\ref{Power})).
Different lines correspond to different ion species (see legend).
Here we assume Gaussian beams with $w_0$ = 20 $\mu$m, and a Rabi
frequency $\Omega_\text{R}/2\pi$ = 0.25 MHz ($\tau_{\pi}=1$
$\mu$sec).} \label{Power_vs_error}
\end{figure}

Figure \ref{Power_vs_error} shows the laser power needed per Raman
beam for a given error due to spontaneous Raman scattering into
the $\text{S}_{1/2}$ manifold. Here we assume
$\Omega_\text{R}/2\pi$ = 0.25 MHz ($\tau_{\pi}=1$ $\mu$sec), and
$w_{0}$ = 20 $\mu$m. Different lines correspond to the different
ion species listed on the figure legend. Table \ref{Table1} lists
the error in a single ion-qubit gate due to Raman scattering back
into the $\text{S}_{1/2}$ manifold for the same parameters as in
Fig. \ref{Power_vs_error}, and assuming $10$ mW in each of the
Raman beams. The power $\mathcal{P}_0$ needed in each of the gate
beams for $\epsilon_\text{S}=10^{-4}$ is also listed in the table.
As can be seen, ions with shorter transition wavelength require a
larger detuning and accordingly higher laser power to maintain a
low gate error. For most ions, a few milliwatts of laser power is
enough to reduce the gate error to below $10^{-4}$.

Equation (\ref{Raman_prob1}) can be solved to give the required
detuning values for a given $\epsilon_\text{S}$. The number of
such detuning values comes from the number of crossings of a
horizontal line, set at the desired error level, with the solid
curve in Fig. \ref{P_Raman_vs_Detuning}. When $\epsilon_\text{S}$
is higher than the minimum $P_{\text{Raman}}$ inside the
fine-structure manifold, i.e.
$\epsilon_\text{S}>8\pi\gamma/3\omega_\text{f}$ (see Fig.
\ref{P_Raman_vs_Detuning}), then four different detuning values
yield the same $\epsilon_\text{S}$, two outside and two inside the
fine-structure manifold. When $\epsilon_\text{S}$ is lower than
this value, only two detuning values, both of which are outside
the fine-structure manifold, yield $\epsilon_\text{S}$. Those
detuning values $\Delta_0$ that are below the $S_{1/2} \rightarrow
P_{1/2}$ transition ($\Delta<0$) and correspond to
$\epsilon_\text{S}=10^{-4}$ are listed in Table \ref{Table1} for
different ions. For most ions $\Delta_0$ is in the few hundred
gigahertz range, and its magnitude is much smaller than
$\omega_\text{f}$.

We now consider the error for various ions caused by Raman
scattering into low lying D levels, $\epsilon_\text{D}$. As
transitions between levels in the S and the D manifolds do not
necessarily involve electron spin rotations, the error suppression
discussed preceding Eq. (\ref{Raman_prob1}) will not occur.
Instead, Raman scattering rate into the D levels will be given by
the total scattering rate times a fixed branching ratio $f$. When
driven resonantly the $\text{P}_{1/2}$ and $\text{P}_{3/2}$ levels
decay to the D manifold with different (but often similar)
branching ratios. Here we assume that for a detuning large
compared to the fine-structure splitting, the branching ratio is
essentially independent of the laser detuning and is given by the
average of the two resonant branching ratios \cite{D1 remark, D2
remark}. Table \ref{Table0} lists $f$, for various ion species,
obtained from \cite{f1,f2,f3,f4,f5}. The error due to Raman
scattering into D levels is given by
\begin{equation}
\epsilon_\text{D}=fP_{\text{total}}. \label{D error}
\end{equation}
Using Eq. (\ref{Raman_prob1}) we can write the ratio of the errors
due to Raman scattering into the different manifolds
\begin{equation}
\frac{\epsilon_\text{D}}{\epsilon_\text{S}}=\frac{3f}{2}\left(\frac{2\Delta^2
+ (\Delta - \omega_\text{f})^2}{\omega_\text{f}^2}\right).
\label{StoD error_ratio}
\end{equation}
For $|\Delta|<\omega_\text{f}$, Raman scattering error is
dominated by scattering back into the $\text{S}_{1/2}$ levels. For
ions with $1/f\gg 1$ the two errors become comparable at a
detuning $\Delta \simeq \sqrt{2}\omega_\text{f}/3\sqrt{f}$. When
the detuning becomes large compared to the fine-structure
splitting, scattering into low-lying D levels dominates. For most
ions considered here, $|\Delta_0|<\omega_\text{f}$ and (perhaps
with the exception of $^{137}$Ba$^+$) $\epsilon_\text{S}$ is the
more dominant source of error. The ratios
$\epsilon_\text{D}/\epsilon_\text{S}$ when
$\epsilon_\text{S}=10^{-4}$ (i.e. $\Delta = \Delta_0$) are given
in Table \ref{Table1} for different ions.

Due to the asymptotic value of the total scattering rate in the
$|\Delta|\gg\omega_\text{f}$ limit (Eq. (\ref{total_limit})),
$\epsilon_\text{D}$ has an asymptotic value which gives a lower
bound to the Raman scattering error
\begin{equation}
\epsilon_{\text{D}\infty}=\frac{3\pi\gamma f}{\omega_\text{f}}.
\label{D errorLimit}
\end{equation}
Table \ref{Table1} lists $\epsilon_{\text{D}\infty}$ for various
ions. For all ion species considered this value is below the
assumed estimates for the fault tolerance threshold.

\begin{table}[tbp]
\begin{center}
  \begin{tabular}{|l|c|c|c|c|c|}
  \hline
  Ion&$\epsilon_\text{S}/10^{-4}$&$\mathcal{P}_0$(mW)&$\Delta_0/2\pi$(GHz)&$\epsilon_\text{D}$/$\epsilon_\text{S}$&$\epsilon_{\text{D}\infty}/10^{-4}$\\
  \hline
  \hline
  $^9$Be$^+$&$0.34$&3.4&-203&N.A.&N.A.\\
  \hline
  $^{25}$Mg$^+$&$0.47$&$4.7$&$-691$&N.A.&N.A.\\
  \hline
  $^{43}$Ca$^+$&$0.17$&$1.7$&$-442$&$0.10$&$0.019$\\
  \hline
  $^{67}$Zn$^+$&$1.23$&$12.3$&$-1247$&N.A.&N.A.\\
  \hline
  $^{87}$Sr$^+$&$0.15$&$1.5$&$-442$&$0.11$&$0.006$\\
  \hline
  $^{111}$Cd$^+$&$1.05$&$10.5$&$-1043$&N.A.&N.A.\\
  \hline
  $^{137}$Ba$^+$&$0.11$&$1.1$&$-418$&$0.51$&$0.012$\\
  \hline
  $^{171}$Yb$^+$&$0.2$&$2$&$-411$&$0.005$&$0.00006$\\
  \hline
  $^{199}$Hg$^+$&$2.3$&$23$&$-1141$&$0.002$&$0.00003$\\
  \hline
  \end{tabular}
\end{center}
\caption{A list of errors in a single-qubit gate ($\pi$ rotation)
due to spontaneous photon scattering. The error due to Raman
scattering back into the $\text{S}_{1/2}$ manifold
$\epsilon_\text{S}$ is calculated with the same parameters as Fig.
\ref{Power_vs_error}: Gaussian beams with $w_0$ = 20 $\mu$m, a
single ion Rabi frequency $\Omega_\text{R}/2\pi$ = 0.25 MHz
($\tau_{\pi}=1$ $\mu$sec), and $10$ mW in each of the Raman beams.
$\mathcal{P}_0$ is the power (in milliwatts) needed in each of the
beams, and $\Delta_0/2\pi$ is the detuning (in gigahertz) for
$\epsilon_\text{S}=10^{-4}$. The ratio between errors due to Raman
scattering to the D and S manifolds,
$\epsilon_\text{D}$/$\epsilon_\text{S}$, is given when
$\epsilon_\text{S}=10^{-4}$.  The asymptotic value of
$\epsilon_\text{D}$ in the $|\Delta|\gg\omega_\text{f}$ limit is
$\epsilon_{\text{D}\infty}$.}\label{Table1}
\end{table}

\subsection{Two-qubit gate}
A universal quantum gate set is complete with the addition of
two-qubit entangling gates. During the last few years there have
been several proposals and realizations of two ion-qubit gates
\cite{CiracZoller, SMgate, Solano, Milburn, SackettGate,
Didi_gate, Innsbruck_gate, Michigan_gate, Oxford_gate}. Here we
focus on gates that use spin-dependent forces in order to imprint
a geometric phase on certain collective spin states \cite{SMgate,
Solano, Milburn, SackettGate, Didi_gate, Michigan_gate,
Oxford_gate}. We examine only gates that are implemented with a
continuous non resonant pulse rather than those using multiple
short pulses \cite{pulsed gate}. Again, to compare different ion
species we examine ion-qubits that are encoded into hyperfine
clock states. When the laser detuning is large compared to the
hyperfine splitting, the differential light force between clock
levels is negligible \cite{HighField, PJLee}. However, a phase
gate can be applied between spin states in the rotated basis
(superpositions of clock states that lie on the equatorial plane
of the Bloch sphere) \cite{SMgate, Solano}. In this scheme the
ions traverse a trajectory in phase space that is conditioned on
their mutual spin state (in the rotated basis
\cite{Michigan_gate}). The phase the ions acquire is proportional
to the total area encircled in phase space. This geometric phase
gate was demonstrated in \cite{SackettGate} and was realized on
clock states in \cite{Michigan_gate}.

This form of phase gate is implemented with two different Raman
fields that are slightly off-resonance with upper and lower
motional sidebands of the spin-flip transition. For simplicity, we
assume here that the gate is driven by two independent pairs of
Raman beams, i.e. a total of four beams. Most experimental
implementations of this phase gate thus far have used a three-beam
geometry \cite{SackettGate, Michigan_gate}. It is straightforward
to generalize the treatment below to the three-beam case.

Typical conditions for the gate are such that the angular
frequency difference between the beams is
$(\omega_0+\omega_{\text{trap}}-\delta)$ in one Raman pair and
$(\omega_0-\omega_{\text{trap}}+\delta)$ in the other. Here
$\omega_{\text{trap}}$ is the angular frequency of the normal mode
that the gate excites and $\delta$ is the Raman field detuning
from that motional sideband. Under these conditions the ions will
traverse $K$ full circles in phase space for a gate duration of
$\tau_{\text{gate}}=2\pi K/\delta$. Typically $|\delta|$ is chosen
to be much smaller than $\omega_{\text{trap}}$ to avoid coupling
to the pure spin flip (``carrier'') transition or the motional
``spectator'' mode, and $K$ is usually chosen to be 1 to minimize
the gate time.

\begin{figure}[tbp]
\begin{center}
\includegraphics[width=8.6cm]{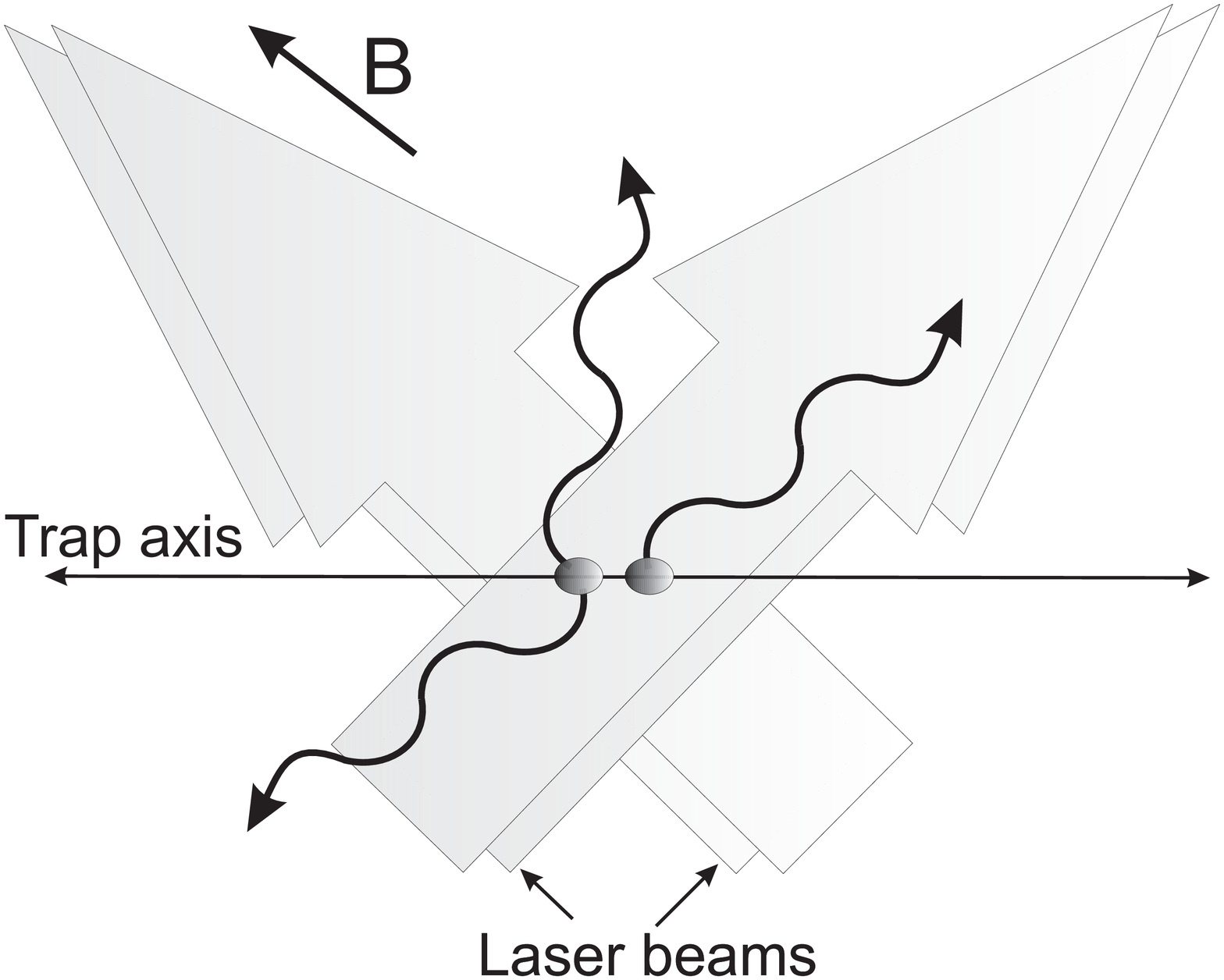}
\end{center}
\caption{Schematic of Raman laser beam geometry assumed for the
two-qubit phase gate. The gate is driven by two Raman fields, each
generated by a Raman beam pair. Each pair consists of two
perpendicular beams of different frequencies that intersect at the
position of the ions such that the difference in their wave vector
lies parallel to the trap axis. One beam of each pair is parallel
to the magnetic field which sets the quantization direction. The
beams' polarizations in each pair are assumed to be linear,
perpendicular to each other and to the magnetic field. Wavy arrows
illustrate examples of photon scattering directions.}
\label{Beam_Layout}
\end{figure}

Figure \ref{Beam_Layout} depicts the assumed geometry of the laser
beams. The two beams comprising each Raman pair intersect at right
angles at the position of the ions, such that the difference in
their wave vectors is parallel to the trap axis. With this choice,
the Raman fields couple, to a very high degree, only to the motion
along the trap axis. The beam polarizations are assumed to be
linear, perpendicular to each other and to the magnetic field
axis. The beams' relative frequencies can be arranged such that
the final state is insensitive to the optical phase at the ions'
position \cite{PJLee}. Generalizing this treatment to other Raman
beam geometries is straightforward.

As in the single-qubit case, Raman scattering will project one of
the ion qubits into one of its states below the P manifold. In the
appropriate basis the ideal gate operation is represented by
\cite{Didi_gate, Michigan_gate}

\begin{equation}
\hat{U}=\begin{pmatrix}1&0&0&0\\0&e^{i\phi}&0&0\\0&0&e^{i\phi}&0\\0&0&0&1\end{pmatrix},
\label{Ideal gate matrix}
\end{equation}
where typically $\phi=\pi/2$. The output state is $|\Psi\rangle =
\hat{U}|\Psi_{\text{init}}\rangle$. The density matrix following
the erroneous gate will be of the form
\begin{eqnarray}
&& \hat{\rho}_{\text{final}} = (1 -
2P_{\text{Raman-gate}})|\Psi\rangle\langle\Psi|+\hat\rho_{\epsilon},
\label{RhoFinal}
\end{eqnarray}
where $P_{\text{Raman-gate}}$ is the probability that one of the
ions scattered a Raman photon during the gate \cite{No Correlation
Remark, multiple scattering remark}. For a gate consisting of $K$
circles, the Raman detuning $\delta$ is chosen such that the gate
time is given by \cite{SMgate}
\begin{equation}
\tau_{\text{gate}} =
\frac{\pi}{2|\Omega_\text{R}|\eta}\sqrt{K}=\tau_{\pi}\frac{\sqrt{K}}{\eta}.
\label{Gate time}
\end{equation}
Here $\eta = \Delta k z_0$ is the Lamb-Dicke parameter, where
$\Delta k=\sqrt{2}k_L$ is the wave vector difference between the
two beams that drive the gate (for the particular geometry of Fig.
\ref{Beam_Layout}), and $k_L$ is the laser beam wave vector
magnitude. The root mean-square of the spatial spread of the
ground state wave function of one ion for the normal mode that the
gate excites is
\begin{equation}
z_0 = \sqrt{\hbar/4M\omega_{\text{trap}}}, \label{z0}
\end{equation}
where $M$ is the mass of an individual ion. The single ion carrier
Rabi frequency $\Omega_\text{R}$ is given in Eq. (\ref{Rabi}).

Using similar considerations to those used in the single ion gate
and, as before, neglecting any (positive) contribution of
$\hat{\rho}_\epsilon$ to the fidelity, we can set an upper bound
for the power needed for a certain error in the gate due to Raman
photon scattering into $\text{S}_{1/2}$ levels
\begin{equation}
\mathcal{P}=\frac{2\pi}{3\epsilon_\text{S}}(\frac{2\pi
w_{0}}{\lambda_{3/2}})^2\hbar\omega_{3/2}|\Omega_\text{R}|\frac{4\sqrt{K}}{\eta}.
\label{Power vs. error Gate}
\end{equation}
This required gate power is $4\sqrt{K}/\eta$ times larger than
that needed for the same error in the single ion-qubit $\pi$
rotation given in Eq. (\ref{Power}). The factor of $\sqrt{K}/\eta$
is due to the longer two ion-qubit gate duration compared to
single qubit rotations, and the factor of $4$ is due to the
presence of two ions and the pair of required Raman fields.

When comparing different ion species we can fix different
parameters, depending on experimental constraints or requirements.
For example, here we choose as fixed parameters the beam waist
$w_{0}$ (for the reasons given above), the gate time
$\tau_{\text{gate}}$ (assuming a certain computation speed is
desired), and the mode frequency $\omega_{\text{trap}}$ (which
sets the time scale for various gates). With these choices,
heavier ions pay the price of smaller $\eta$ and therefore higher
power requirements per given gate time and error. A different
approach would be to choose $\eta$ fixed, in which case heavier
ions will need a lower $\omega_{\text{trap}}$. For a fixed gate
time a lower $\omega_{\text{trap}}$ leads to stronger,
off-resonant, coupling of the Raman fields to the carrier or the
other motional ``spectator'' mode, and, therefore, to a larger
error due to this coupling \cite{SMgateThermal, heavier ions}. The
Lamb-Dicke parameters for the different ions for
$\omega_{\text{trap}}/2\pi$ = 5 MHz are listed in Table
\ref{Table2}.

\begin{figure}[tbp]
\begin{center}
\includegraphics[width=8.6cm]{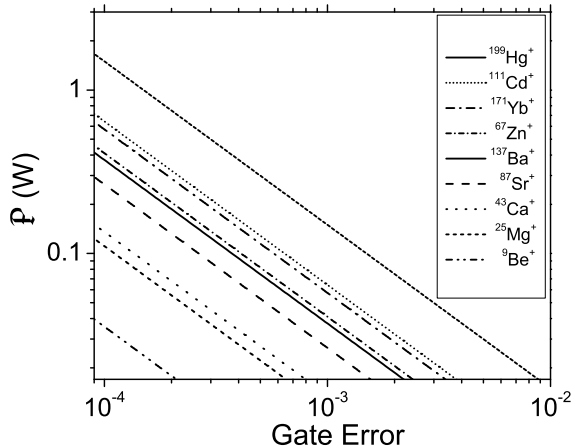}
\end{center}
\caption{Laser power in each of the Raman beams vs. the error in a
two ion entangling gate due to Raman scattering back into the
$\mathcal{S}_{1/2}$ manifold (obtained using Eq. (\ref{Power vs.
error Gate})). Different lines correspond to different ion species
(see legend). Here we assume Gaussian beams with $w_0$ = 20
$\mu$m, a gate time $\tau_{\text{gate}}$ = 10 $\mu$s,
$\omega_{\text{trap}}/2\pi$ = 5 MHz, and a single circle in phase
space ($K$ = 1).} \label{Power_vs_error2}
\end{figure}

With the choice where $w_0$, $\tau_{\text{gate}}$, and
$\omega_{\text{trap}}$ are fixed, we can write
\begin{equation}
\mathcal{P}=\frac{8\pi^2}{3\epsilon_\text{S}}\frac{K}{\tau_{\text{gate}}}w_{0}^2\omega_{3/2}M\omega_{\text{trap}}.
\label{Power vs. error Gate}
\end{equation}
Or equivalently,
\begin{equation}
\epsilon_\text{S}=\frac{8\pi^2}{3\mathcal{P}}\frac{K}{\tau_{\text{gate}}}w_{0}^2\omega_{3/2}M\omega_{\text{trap}}.
\label{error Gate vs. Power}
\end{equation}
Figure \ref{Power_vs_error2} shows the power needed vs. error in a
two-qubit gate due to Raman scattering back into the
$\text{S}_{1/2}$ manifold for $w_{0}$ = 20 $\mu$m,
$\tau_{\text{gate}}$ = 10 $\mu$s, $\omega_{\text{trap}}/2\pi$ = 5
MHz and $K$ = 1. Table \ref{Table2} lists $\epsilon_\text{S}$ for
the various ion species and the same laser parameters as Fig.
\ref{Power_vs_error2}, assuming a power of $10$ mW is used in each
of the four Raman beams. Alternatively, the power $\mathcal{P}_0$
and the detuning $\Delta_0$ needed in each of the gate beams for
$\epsilon_\text{S}$ = $10^{-4}$ are also listed in the table.
Here, heavier ions need a larger detuning and correspondingly
higher laser power to maintain a low gate error. For most ion
species, hundreds of milliwatts of laser power per beam and a
detuning comparable or even larger than $\omega_\text{f}$ are
needed to reduce the gate error to the $10^{-4}$ level.

As in the one-ion gate, Raman scattering into low lying D levels
will add to the gate error. This error is given by
\begin{equation}
\epsilon_\text{D}=2fP_{\text{total-gate}}=\frac{4\sqrt{K}}{\eta}fP_{\text{total}}.
\label{D error gate}
\end{equation}
Here $P_{\text{total-gate}}$ is the probability that one of the
ions scattered a photon during the two-qubit gate, and
$P_{\text{total}}$ is the one-qubit gate scattering probability
given in Eq. (\ref{photon per pulse}). Since both the Raman and
the total scattering probabilities increase by the same factor as
compared to the one-qubit gate, the ratio of the two errors
$\epsilon_\text{D}/\epsilon_\text{S}$ will remain the same as
given by Eq. (\ref{StoD error_ratio}). Table \ref{Table2} lists
$\epsilon_\text{D}/\epsilon_\text{S}$ for the different ions when
$\epsilon_\text{S}$ = $10^{-4}$. Notice that for
$\epsilon_\text{S}$ = $10^{-4}$ some ions require $|\Delta_0|
\gtrsim \sqrt{2}\omega_\text{f}/3\sqrt{f}$. For those ions
$\epsilon_\text{D}$ is no longer negligible compared to
$\epsilon_\text{S}$.

Scattering into a low lying D level will, again, set a lower bound
on the total error. Combining Equations (\ref{D error gate}) and
(\ref{total_limit}) we find this lower bound to be
\begin{equation}
\epsilon_{\text{D}\infty}=\frac{3\pi\gamma
f}{\omega_\text{f}}\frac{4\sqrt{K}}{\eta}. \label{DerrorGate}
\end{equation}
Table \ref{Table2} lists $\epsilon_{\text{D}\infty}$ for the
different ion species.

\begin{table}[tbp]
\begin{center}
  \begin{tabular}{|l|c|c|c|c|c|c|}
  \hline
  Ion&$\epsilon_\text{S}/10^{-4}$&$\mathcal{P}_0$(mW)&$\Delta_0/2\pi$(THz)&$\epsilon_\text{D}/\epsilon_\text{S}$&$\epsilon_{\text{D}\infty}/10^{-4}$&$\eta$\\
  \hline
  \hline
  $^9$Be$^+$&3.6&36&-1.20&N.A.&N.A.&0.194\\
  \hline
  $^{25}$Mg$^+$&11.1&111&-7.28&N.A.&N.A.&0.130\\
  \hline
  $^{43}$Ca$^+$&13.6&136&-10.42&1.01&1.06&0.071\\
  \hline
  $^{67}$Zn$^+$&41.1&411&-24.96&N.A.&N.A.&0.11\\
  \hline
  $^{87}$Sr$^+$&26.5&265&-20.34&0.52&0.50&0.048\\
  \hline
  $^{111}$Cd$^+$&64.3&643&-35.44&N.A.&N.A.&0.081\\
  \hline
  $^{137}$Ba$^+$&37.4&374&-30.67&1.65&1.46&0.034\\
  \hline
  $^{171}$Yb$^+$&57.5&575&-32.89&0.01&0.007&0.038\\
  \hline
  $^{199}$Hg$^+$&149.7&1497&-49.52&0.003&0.001&0.078\\
  \hline
  \end{tabular}
\end{center}
\caption{A list of different errors in a two-qubit phase gate due
to spontaneous photon scattering. The error due to Raman
scattering back into the $\mathcal{S}_{1/2}$ manifold,
$\epsilon_\text{S}$, is calculated assuming Gaussian beams with
$w_0$=20 $\mu$m, a gate time $\tau_{\text{gate}}$=10 $\mu$s,
$\omega_{\text{trap}}/2\pi=$5 MHz, a single circle in phase space
($K$=1), and $10$ mW in each of the four Raman beams.
$\mathcal{P}_0$ is the power in milliwatts needed in each of the
beams, and $\Delta_0/2\pi$ is the detuning in gigahertz for
$\epsilon_\text{S}=10^{-4}$. The ratio between errors due to Raman
scattering to the D and S manifolds,
$\epsilon_\text{D}$/$\epsilon_\text{S}$, is given when
$\epsilon_\text{S}=10^{-4}$.  The asymptotic value of
$\epsilon_\text{D}$ in the $|\Delta|\gg\omega_\text{f}$ limit is
$\epsilon_{\text{D}\infty}$. The Lamb-Dicke parameter $\eta$ for
the above trap frequency is also listed for different ions.}
\label{Table2}
\end{table}

\section{RAYLEIGH SCATTERING ERROR}\label{section3}
Since Rayleigh photon scattering is elastic, no energy or angular
momentum is transferred  between the photons' and the ions'
internal degrees of freedom. Therefore, these degrees of freedom
remain uncorrelated. Rayleigh scattering does not necessarily lead
to direct spin decoherence.

In situations where Rayleigh scattering rates from the two
ion-qubit states are different, Rayleigh scattering of photons
will eventually measure the qubit state and lead to decoherence.
In fact, the most common ion-qubit detection method relies on
state selective Rayleigh scattering of photons on a cycling
transition. In most Raman gates, however, the laser is detuned
from resonance by much more than the qubit levels' energy
separation, $\Delta \gg \omega_0$, and the laser polarization is
typically linear to suppress differential Stark shifts. Under
these conditions Rayleigh scattering rates from the two qubit
levels are almost identical and the error due to the rate
difference is negligible. For the clock transition qubit states
considered here, Rayleigh scattering rates are almost identical,
regardless of the laser polarization (for more details on this
error see subsection \ref{other errors}).

The main effect of Rayleigh scattering is the momentum recoil it
imparts to the ion-qubit. For a single-qubit gate, the Raman beams
are usually arranged in a co-propagating geometry. In this
configuration, since the Lamb-Dicke parameter is very small
($\eta\simeq(\omega_0/\omega_{3/2}) k_L z_0$), the effect of ion
motion is negligible on the gate operation. Therefore, Rayleigh
scattering has a negligible effect on single-qubit gates.

In the two-qubit phase gate, a mode of motion is excited that is
entangled with the two-ion collective spin state. In this case,
recoil from photon scattering perturbs the ion's motion through
phase space and contributes to the gate error. The ion-qubit
trajectory is distorted in two ways. The larger distortion arises
from the direct recoil momentum displacement. A second, much
smaller, distortion arises from the contribution of the recoil to
the appearance of nonlinearities in the gate operation due to
deviations from the Lamb-Dicke regime (see subsection \ref{other
errors}).

In subsection \ref{recoil_displacement_error} we calculate the
error in a two-qubit gate due to direct recoil phase-space
displacement. In subsection \ref{other errors} we elaborate on the
two other sources of error mentioned above, namely errors due to
uneven Rayleigh scattering rates and errors due to deviations from
the Lamb-Dicke regime.

\subsection{Rayleigh scattering recoil error}\label{recoil_displacement_error}
In elastic Rayleigh scattering, energy and momentum are not
exchanged between the ions' and the photons' internal degrees of
freedom. However, momentum and energy are exchanged between the
photons' and the ions' external degrees of freedom. The scattered
photon direction will be different from that of the laser beam,
causing the ion to recoil. This recoil acts on the ion-qubit as a
phase-space momentum displacement, distorting the ion's trajectory
through phase space and causing an error in the entangling-gate
phase. Note that the momentum imparted to the ion (and therefore
the deviation from the desired gate phase) and the momentum that
is carried by the scattered photon (namely its scattering
direction) are correlated. From this point of view the gate
infidelity again arises due to the entanglement between the
scattered photons' and ions' (this time external) degrees of
freedom (for more details on this point of view see Appendix
\ref{appendix1}).

In the Lamb-Dicke regime (for a thermal state $\eta\sqrt{2 \bar{n}
+1} \ll 1$, where $\bar{n}$ is the average mode population), the
gate operation can be approximated as a series of finite
displacements in phase space \cite{Didi_gate}

\begin{equation}
\hat{U} =\prod_{k = 1}^{N} \hat{D}(\Delta\alpha_{k}). \label{Ideal
gate as displacements}
\end{equation}

Displacements through phase space, $\Delta\alpha_{k}$, are
conditioned on the joint spin state of the two ions, and depend on
the gate parameters. For certain gate parameters, the displacement
is zero for the two parallel spin states
$|\uparrow\uparrow\rangle$ and $|\downarrow\downarrow\rangle$, and
non zero with opposite sign for the $|\uparrow\downarrow\rangle$
and $|\downarrow\uparrow\rangle$ states, where $\uparrow$ and
$\downarrow$ hereafter refer to the rotated basis rather than the
clock levels. Using the commutation relations between phase space
displacements
\begin{equation}
\hat{D}(\alpha)\hat{D}(\beta)=\hat{D}(\alpha+\beta)e^{i\text{Im}(\alpha\beta^*)},
\label{displacements commutation}
\end{equation}
we can write the gate operation as a single displacement times an
overall phase \cite{Didi_gate}
\begin{equation}
\hat{U} = \hat{D}(\sum_{k =
1}^{N}\Delta\alpha_{k})e^{i\text{Im}(\sum_{j =
2}^{N}\Delta\alpha_{j}\sum_{l = 1}^{j-1}\Delta\alpha_{l})},
\label{Ideal gate as displacements2}
\end{equation}
where $N$ is the total number of infinitesimal displacements. When
the net displacement is zero, i.e. $\sum_{k =
1}^{N}\Delta\alpha_{k}=0$, the two-ion motion returns to its
initial state and spin and motion are disentangled at the end of
the gate. The gate operation on the affected spin states can be
written as
\begin{equation}
\hat{U} = \hat{I}e^{i\phi}. \label{Ideal gate as displacements3}
\end{equation}
The phase acquired,
\begin{equation}
\phi = \text{Im}(\sum_{j = 2}^{N}\Delta\alpha_{j}\sum_{l =
1}^{j-1}\Delta\alpha_{l}), \label{gate_phase}
\end{equation}
is the same for the $|\uparrow\downarrow\rangle$ and
$|\downarrow\uparrow\rangle$ states and proportional to the
encircled phase-space area. Figure \ref{Phase space figure}a shows
an example for the trajectory traversed in phase space during an
ideal gate in a reference frame rotating at the mode frequency
$\omega_{\text{trap}}$.

In the presence of laser light there is a finite probability
$P_{\text{Rayleigh-gate}}$ that a Rayleigh photon will be
scattered by one of the ions during the gate \cite{multiple
scattering remark}.

The probability for a Rayleigh scattering event to occur during a
single-qubit gate is given by the difference between Eq.
(\ref{photon per pulse}) and Eq. (\ref{Raman_prob1})
\begin{equation}
P_{\text{Rayleigh}}=\frac{\pi\gamma}{\omega_\text{f}}\frac{3\Delta^2-2\Delta\omega_\text{f}+\omega_\text{f}^2/3}{|\Delta(\Delta-\omega_\text{f})|}.
\label{Rayleigh_prob}
\end{equation}
Since in the limit of $|\Delta| \gg \omega_\text{f}$ all
scattering events are Rayleigh scattering, $P_{\text{Rayleigh}}$
and $P_{\text{total}}$ have the same asymptotic value
$P_{\infty}$.

Using Eq. (\ref{Rayleigh_prob}) and the factor for the extra
required power in the two-qubit gate
\begin{equation}
P_{\text{Rayleigh-gate}}=\frac{4\sqrt{K}\pi\gamma}{\eta\omega_\text{f}}\left(\frac{3\Delta^2-2\Delta\omega_\text{f}+\omega_\text{f}^2/3}{|\Delta(\Delta-\omega_\text{f})|}\right).
\label{Rayleigh_prob two ion gate}
\end{equation}
The effect on the ion motion would be that of momentum recoil
\begin{equation}
\hat{U}_{\text{recoil}} = e^{i\bf{q}\cdot\hat{\bf{r}}_i}.
\label{recoil_operator1}
\end{equation}
Here $\bf{q}= \bf{k_\text{L}} - \bf{k_{\text{Scat}}}$ is the wave
vector difference between the scattered photon and the laser beam
from which it was scattered, and $\hat{\bf{r}}_i$ is the position
operator of the ion that scattered the photon ($i=1,2$). We will
neglect recoil into directions other than along the trap axis. We
can write the position operator for for ion $i$ along this axis as
\begin{equation}
\hat{\bf{z}}_i = Z_i + z_0(\hat{a}_0 + \hat{a}^\dagger_0) +
z_1(\hat{a}_1 + \hat{a}^\dagger_1), \label{position_operator}
\end{equation}
where $Z_i$ is the equilibrium position of the ion and $z_{0/1}$,
$\hat{a}^\dagger_{0/1}$ and $\hat{a}_{0/1}$ are the root
mean-square of the ground state spatial spread, and the creation
and annihilation operators, respectively, of the two motional
modes $0$ and $1$. We assume the gate is performed by exciting
mode $0$. Recoil into mode $1$ does not distort the gate dynamics
directly. This part of the photon recoil will add minutely to the
gate error through its contribution to the gate nonlinearity
discussed in section \ref{other errors}. We neglect this
contribution for the moment and write the recoil operation as
\begin{equation}
\hat{U}_{\text{recoil}} = e^{iq_z z_0(\hat{a}_0 +
\hat{a}^\dagger_0)} \equiv e^{\beta\hat{a}_0 -
\beta^{\star}\hat{a}^\dagger_0} \equiv \hat{D}(\beta),
\label{recoil_operator2}
\end{equation}
where $q_z$ is the projection of $\bf{q}$ along the trap axis, and
$\beta = i q_z z_0$.

The recoil action on the two-ion crystal is, therefore, a momentum
displacement $\beta$ in phase space. The magnitude of $\beta$
depends on the wave-vector difference between the scattered photon
and the beam from which it was scattered, and the projection of
this momentum difference along the trap axis. Since the same
recoil displacement is applied to all spin states, motion will
still be disentangled from spin at the end of the gate. Errors are
therefore due to the change in area encircled in phase space.
Figure \ref{Phase space figure}b illustrates a gate trajectory
that is distorted due to photon recoil.

\begin{figure}[tbp]
\begin{center}
\includegraphics[width=8.6cm]{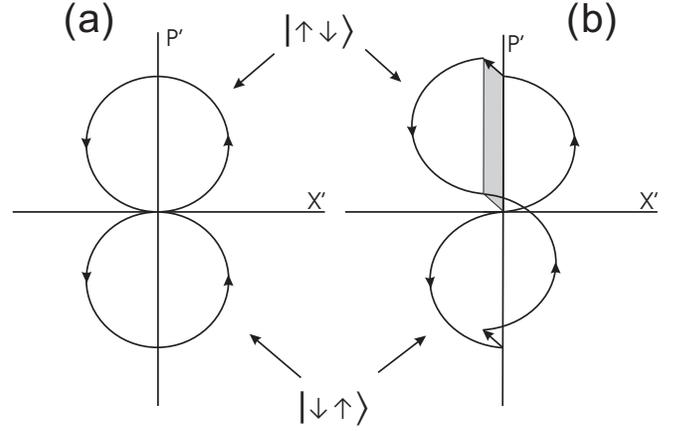}
\end{center}
\caption{Schematics of the trajectories traversed by the ions in
phase-space during the gate. The acquired phase is proportional to
the encircled area. (a) Phase-space trajectory of an ideal gate.
(b) Phase-space trajectory of an erroneous gate where a photon was
scattered during the gate drive. The short straight arrows
represent the recoil displacement. The grey-shaded area is
proportional to the phase error. The area that is added to the
$|\uparrow\downarrow\rangle$ trajectory is subtracted from the
$|\downarrow\uparrow\rangle$ trajectory.} \label{Phase space
figure}
\end{figure}

The erroneous gate can be again written as a sum of displacements,
those due to the gate drive and $\beta$, which occurs at some
random time during the gate. For a particular two-ion spin state,
\begin{equation}
\hat{U}_{\epsilon} =\prod_{k =M+1}^{N}
\hat{D}(\Delta\alpha_{k})\hat{D}(\beta)\prod_{l =1}^{M}
\hat{D}(\Delta\alpha_{l}).\label{erroneous gate as displacements2}
\end{equation}
By use of the commutation relation (\ref{displacements
commutation}), the gate is written as
\begin{equation}
\hat{U}_{\epsilon}
=\hat{D}(\beta)e^{i(\phi+\Delta\phi)},\label{erroneous gate as
displacements3}
\end{equation}
where the phase error $\Delta\phi$ is determined to be
\begin{equation}
\Delta\phi = \text{Im}\left[\beta\sum_{l =1}^{M}\Delta\alpha_{l}^*
+ \sum_{k =M+1}^{N} \Delta\alpha_{k}\beta^*\right].\label{phase
error1}
\end{equation}
Writing displacements as a function of the gate time
\begin{equation}
\Delta\alpha_{k} = \Delta\alpha(t_{k}) \rightarrow
\frac{\partial\alpha(t)}{\partial t}dt,\label{time_displacement}
\end{equation}
we can write the phase error as a function of the gate
displacement before and after the scattering time
$t_{\text{scat}}$
\begin{equation}
\Delta\phi =
\text{Im}\left[\beta\int_{0}^{t_{\text{scat}}}\frac{\partial\alpha^*(t)}{\partial
t}dt +
\beta^*\int_{t_{\text{scat}}}^{t_{\text{gate}}}\frac{\partial\alpha(t)}{\partial
t}dt\right].\label{phase error2}
\end{equation}
Since $\alpha(t_{\text{gate}}) = \alpha(0) = 0$ we get
\begin{equation}
\Delta\phi =
\text{Im}[\beta\alpha^*(t_{\text{scat}})-\beta^*\alpha(t_{\text{scat}})].\label{phase
error2}
\end{equation}
Since the accumulated displacement at the moment the photon was
scattered $\alpha(t_{\text{scat}})$ is of equal magnitude but
opposite sign for the $|\uparrow\downarrow\rangle$ and
$|\downarrow\uparrow\rangle$ states, we get
\begin{equation}
\Delta\phi_{\uparrow\downarrow} =
-\Delta\phi_{\downarrow\uparrow}\equiv\Delta\phi.\label{phase
error3}
\end{equation}

The erroneous gate can be therefore represented by the operator
\begin{equation}
\hat{U}_{\epsilon}=\begin{pmatrix}1&0&0&0\\0&e^{i(\phi+\Delta\phi)}&0&0\\0&0&e^{i(\phi-\Delta\phi)}&0\\0&0&0&1\end{pmatrix}.
\label{erroneous gate matrix}
\end{equation}
The final state fidelity following the erroneous gate depends on
$|\Psi_{\text{init}}\rangle$. For a general initial state
\begin{equation}
|\Psi_{\text{init}}\rangle=\mu|\uparrow\uparrow\rangle+\kappa|\uparrow\downarrow\rangle+\gamma|\downarrow\uparrow\rangle+\delta|\downarrow\downarrow\rangle,
\label{initial state}
\end{equation}
and using Eq. (\ref{fidelity}), we can write for the gate fidelity
\begin{equation}
F=\left||\mu|^2+e^{i\Delta\phi}|\kappa|^2 +
e^{-i\Delta\phi}|\gamma|^2 + |\delta|^2\right|^2, \label{general
gate fidelity}
\end{equation}
The most relevant fidelity for fault tolerance considerations is
that averaged over all possible initial states. The fidelity we
calculate here is that due to the worst-case input state, that is
the input state that minimizes Eq. (\ref{general gate fidelity}).
The worst-case fidelity is clearly smaller than the average
fidelity and therefore gives a conservative estimate to the gate
error \cite{worstcase_vs_average}. The worst-case fidelity for
this kind of error was calculated in \cite{Sasura_Steane}. For
$\Delta\phi/\phi < 1$ it is
\begin{equation}
F = \cos^2\Delta\phi. \label{worst case fidelity1}
\end{equation}
This minimal fidelity is the result of an input state with
$|\kappa^2|=|\gamma^2|=1/2$.

Photon scattering occurs in only a small fraction
$P_{\text{Rayleigh-gate}}$ of the gates. Averaging over all gates
performed, we get
\begin{equation}
F_{\text{gate}} =1-P_{\text{Rayleigh-gate}}(1-\langle
\cos^2\Delta\phi\rangle)\equiv1-\epsilon_\text{R}, \label{worst
case fidelity}
\end{equation}
where the brackets correspond to an average over all possible
$\Delta\phi$'s, i.e. due to different values of $\beta$ and
different scattering times.

To perform the above average we need to explicitly write the
different displacements. In the rotating frame the cumulative gate
displacement after a time $t$ can be written as \cite{Didi_gate,
SMgate}
\begin{equation}
\alpha(t) = \frac{i}{2\sqrt{K}}[e^{-i\delta t}-1]e^{i\Phi_{L} },
\label{acctual displacement}
\end{equation}
where $\Phi_{L}$ is the gate phase, determined by the optical
phase difference between the gate beams at the ion's position. In
the rotating frame, the recoil displacement can be written as
\begin{equation}
\beta = |\beta|e^{i\omega_{\text{trap}}t}. \label{recoil
displacement}
\end{equation}
Substituting equations (\ref{recoil displacement}) and
(\ref{acctual displacement}) into Eq. (\ref{phase error2}), we can
write the averaged term in the fidelity
\begin{widetext}
\begin{eqnarray}
\langle \cos^2\Delta\phi\rangle =
\frac{\delta}{2\pi}\int_{|\beta|=0}^{|\beta|_{\text{max}}}\int_{t=0}^{\frac{2\pi}{\delta}}S(|\beta|)\cos^2\left[\frac{|\beta|}{\sqrt{K}}\left[\cos((\omega_{\text{trap}}+\delta)t-\Phi_{L})-\cos(\omega_{\text{trap}}t-\Phi_{L})\right]\right]dt
d|\beta|. \label{complicated expression of avgcos2}
\end{eqnarray}
\end{widetext} Here we assume that the probability of scattering at
different time intervals during the gate is uniform. The
probability distribution for different recoil displacement
magnitude $|\beta|$ is given by $S(|\beta|)$, which is determined
by the geometry of the Raman beams with respect to the trap axis
and the probability distribution of photon scattering directions.

With the laser beam configuration assumed in
Fig.(\ref{Beam_Layout}), recoil due to photon absorption can be
imparted only at $45^{\circ}$ to the trap axis, whereas recoil due
to photon emission can be imparted in any direction. The maximum
allowed displacement following an absorption-emission cycle is
therefore
\begin{equation}
|\beta|_{\text{max}}=k_L z_0(1+\sqrt{2})/\sqrt{2} =
\frac{\eta}{2}(1+\sqrt{2}). \label{max recoil displacement}
\end{equation}
For proper gate operation we require $\eta \ll 1$; thus we expand
Eq. (\ref{complicated expression of avgcos2}) in powers of
$|\beta|$, including terms to order $|\beta|^2$. Further, since
typically $\delta/\omega_{\text{trap}} \ll 1$, we neglect terms
proportional to $\delta/\omega_{\text{trap}}$. With these
approximations the gate fidelity is independent of $\Phi_{L}$ and
is given by
\begin{equation}
F_{\text{gate}} = 1-P_{\text{Rayleigh-gate}}\frac{\langle
|\beta|^2 \rangle}{2K}. \label{Rayleigh recoil fidelity2}
\end{equation}
Here $\langle |\beta|^2 \rangle$ is the average of the recoil
displacement magnitude squared
\begin{equation}
\langle |\beta|^2 \rangle =
\int_{0}^{|\beta|_{\text{max}}}S(|\beta|)|\beta|^2d|\beta|.
\label{Average kick}
\end{equation}

For the assumed laser beam configuration, the light field
polarization will oscillate rapidly during the gate between
left/right circular and linear at the ions' position. Therefore,
the scattering probability will average to be isotropic. For the
beam configuration illustrated in Fig.\ref{Beam_Layout}, in
spherical-polar coordinates,
\begin{equation}
\frac{\langle |\beta|^2 \rangle}{\eta^2} =\frac{1}{8\pi}
\int_{0}^{2\pi}\int_{0}^{\pi}\left(\frac{1}{\sqrt{2}}+\cos\theta\right)^2\sin\theta
d\theta d\phi. \label{Average kick2}
\end{equation}
Here the $\hat{z}$ ($\theta = 0$) direction is chosen parallel to
the trap axis. We find
\begin{equation}
\langle |\beta|^2 \rangle = \frac{5}{12}\eta^2. \label{beta square
average}
\end{equation}
Combining Eqs. (\ref{worst case fidelity}), (\ref{Rayleigh recoil
fidelity2}), (\ref{beta square average}) and (\ref{Rayleigh_prob
two ion gate}), the Rayleigh error can be written as
\begin{equation}
\epsilon_\text{R}=\frac{5\eta\pi\gamma}{6\sqrt{K}\omega_\text{f}}\left(\frac{3\Delta^2-2\Delta\omega_\text{f}+\omega_\text{f}^2/3}{|\Delta(\Delta-\omega_\text{f})|}\right).
\label{Rayleigh_error vs. detuning}
\end{equation}
The ratio of $\epsilon_\text{R}$ to $\epsilon_\text{S}$ is
therefore
\begin{equation}
\frac{\epsilon_\text{R}}{\epsilon_\text{S}}=\frac{5\eta^2}{16K}\left(\frac{3\Delta^2-2\Delta\omega_\text{f}+\omega_\text{f}^2/3}{\omega_\text{f}^2}\right).
\label{Rayleigh_error to Raman error ratio}
\end{equation}
Table \ref{Table3} lists $\epsilon_\text{R}/\epsilon_\text{S}$ for
different ion species for $\omega_{\text{trap}}/2\pi$=5 MHz, and a
single-circle ($K$=1) gate, at a detuning where
$\epsilon_\text{S}=10^{-4}$. For most ions considered here, the gate
error is still dominated by Raman scattering at this level.

For laser beam detuning large compared to the excited state
fine-structure splitting ($\Delta \gg \omega_\text{f}$) scattering
is dominated by Rayleigh events. In this limit, the probability
that a Rayleigh photon will be scattered during the gate is
\begin{equation}
P_{\text{Rayleigh-gate}\infty} = \frac{4\sqrt{K}P_{\infty}}{\eta}.
\label{scattering prob gate}
\end{equation}
The recoil error in a two-qubit gate is therefore asymptotically
bound by
\begin{equation}
\epsilon_{\text{R}\infty} =
\frac{5\pi\eta}{2\sqrt{K}}\frac{\gamma}{\omega_\text{f}}
=\frac{5\pi^2\gamma}{\omega_\text{f}\lambda_{3/2}}\sqrt{\frac{\hbar}{2M\omega_{\text{trap}}K}}.
\label{recoil error}
\end{equation}

Table \ref{Table3} lists $\epsilon_{\text{R}\infty}$ for different
ion species for $\omega_{\text{trap}}/2\pi$=5 MHz and a
single-circle ($K$=1) gate. With the exception of $^9$Be$^+$, the
error due to photon recoil in a two ion-qubit gate is below
$10^{-4}$. For this error heavier ions benefit due to their smaller
recoil.

It is possible to reduce $\epsilon_\text{R}$ by choosing a smaller
$\eta/\sqrt{K}$ (by increasing the trap frequency and/or performing
multiple-circles gates). Correspondingly more laser power,
proportional to $K/\eta^2$, is then required in order not to
increase the Raman scattering error $\epsilon_\text{S}$ or reduce
the gate speed.

\begin{table}[tbp]
\begin{center}
  \begin{tabular}{|l|c|c|}
  \hline
  Ion&$\epsilon_\text{R}/\epsilon_\text{S}$&$\epsilon_{\text{R}\infty}/10^{-4}$\\
  \hline
  \hline
  $^9$Be$^+$&1.442&1.51\\
  \hline
  $^{25}$Mg$^+$&0.142&0.154\\
  \hline
  $^{43}$Ca$^+$&0.0168&0.0187\\
  \hline
  $^{67}$Zn$^+$&0.0172&0.0193\\
  \hline
  $^{87}$Sr$^+$&0.0030&0.0034\\
  \hline
  $^{111}$Cd$^+$&0.0040&0.0044\\
  \hline
  $^{137}$Ba$^+$&0.0010&0.0011\\
  \hline
  $^{171}$Yb$^+$&0.0006&0.0006\\
  \hline
  $^{199}$Hg$^+$&0.0015&0.0012\\
  \hline
  \end{tabular}
\end{center}
\caption{The ratio $\epsilon_\text{R}$/$\epsilon_\text{S}$ of
errors due to Rayleigh scattering recoil and Raman scattering to
the S manifold, is given when $\epsilon_\text{S}=10^{-4}$.  The
asymptotic value of $\epsilon_\text{R}$ in the
$|\Delta|\gg\omega_\text{f}$ limit is $\epsilon_{\text{R}\infty}$.
Both are calculated assuming $\omega_{\text{trap}}/2\pi = 5$ MHz
and $K=1$.} \label{Table3}
\end{table}

\subsection{Other Rayleigh scattering errors}\label{other errors}
In section \ref{recoil_displacement_error} we calculated the error
due to the recoil imparted to the ion-qubit during Rayleigh photon
scattering. As noted, Rayleigh scattering of photons adds two more
contributions to the gate error.

The first contribution $\epsilon_{\delta}$ is due to a difference in
the Rayleigh scattering rates from the two ground state levels.
Assume that the Rayleigh scattering rate is
$\Gamma_{\text{Rayleigh}}$ from one qubit level and
$\Gamma_{\text{Rayleigh}} + d\Gamma$ from the other. A measurement
of the qubit level will be conclusive once the difference in the
number of photons that are scattered is larger than the standard
deviation of the number of photons scattered from each level. Since
the number of photons that are scattered follows a Poisson
distribution, a measurement will occur after photons have been
scattered for a time $t$ such that
\begin{equation}
d\Gamma t > \sqrt{\Gamma_{\text{Rayleigh}}t}. \label{scattering
difference}
\end{equation}
The error rate is, therefore, given by
$d\Gamma^2/\Gamma_{\text{Rayleigh}}$, and the error during a gate
by
\begin{equation}
\epsilon_{\delta}=(d\Gamma /\Gamma_{\text{Rayleigh}})^2
P_{\text{Rayleigh}}. \label{scattering difference error}
\end{equation}

For the above gate parameters, the difference in scattering rates
is due solely to the difference in detuning between the two qubit
levels. For $\Delta \gg \omega_0$, which is typically required to
reduce other scattering errors to low levels, this error can be
approximated by $\epsilon_{\delta}\simeq(\omega_0/\Delta)^2
P_{\text{Rayleigh}}$. The contribution of $\epsilon_{\delta}$ to
the total error at a realistic laser detuning is very small. We
performed an accurate calculation of the difference in the
scattering rates between the two clock levels, and verified that
indeed, for all ion species and laser detunings discussed above,
$\epsilon_{\delta}$ is negligible.

A second source of Rayleigh scattering error is through the
contribution of the recoil momentum displacement to
non-linearities in the gate evolution. The gate error due to
non-linearities was calculated in \cite{SMgateThermal} and in the
present context is proportional to $P_{\text{Rayleigh-gate}}\eta^4
\text{Var}(n)$, where $\text{Var}(n)$ is the variance of the
motional modes distribution due to recoil. The recoil momentum
displacement magnitude is of order $\eta$. Therefore, starting
from the ground state and following a single scattering event,
$\text{Var}(n)\simeq \eta$. The gate error due to this effect will
be proportional to $P_{\text{Rayleigh-gate}}\eta^5\propto\eta^4$,
and significantly smaller than other scattering errors discussed
above.

\section{DISCUSSION}\label{section4}
We have calculated the errors due to photon scattering in a
single-pulse single-qubit gates and two-qubit phase-gates
implemented with stimulated Raman transitions. These errors
present a fundamental limit to the gate fidelity in trapped-ion
QIP experiments that use these kinds of gates, and should be a
significant factor when choosing a specific ion as a quantum
information carrier for fault-tolerant quantum computing schemes.

Three main errors occur from spontaneously scattering photons
during a gate. Two errors are due to Raman scattering either back
into the $\text{S}_{1/2}$ manifold ($\epsilon_\text{S}$) or to
low-lying D levels ($\epsilon_\text{D}$). The third error is due
to the Rayleigh scattering recoil during a two-qubit gate
($\epsilon_\text{R}$). For most ions currently considered for QIP
experiments, the dominant error for realistic laser parameters is
$\epsilon_\text{S}$. This error can be typically reduced to below
current estimates for the fault-tolerance threshold with the use
of relatively high (but probably attainable) laser power. This
makes the availability of high power laser sources at the relevant
wavelength important. This error is also reduced for ions with a
relatively large $S \rightarrow P$ transition wavelength.

Among those three errors, only $\epsilon_D$ cannot be reduced
below a certain value by the use of higher laser intensity.
Therefore, it may eventually be advantageous to choose ions that
do not have low lying D levels. However, for most ions considered
here, $\epsilon_\text{D}$ is still small at realistic laser
parameters compared to the other Raman error $\epsilon_\text{S}$.

The masses of the different ions play an interesting role in the
scattering error. Since it is harder to transfer momentum to
heavier ions, they suffer from a larger $\epsilon_\text{S}$ in
two-qubit gates (or alternatively from the need for more laser
power for a given gate speed and error level). For the same
reason, lighter ions suffer from a larger Rayleigh recoil error
$\epsilon_\text{R}$. Examining Eqs. (\ref{error Gate vs. Power})
and (\ref{recoil error}), the Raman error scales as
$w_{0}^2M\omega_{\text{trap}}$ and the Rayleigh error scales as
$1/\omega_\text{f}\sqrt{M\omega_{\text{trap}}}$ (neglecting
differences in wavelength and natural linewidths). Since both
$w_{0}$ and $\omega_\text{f}$ are generally larger for heavier
ions, the Raman error is larger and Rayleigh error is smaller for
heavier ions by more than is indicated by the kinetic argument
above. Since for most ions $\epsilon_\text{S}$ is the dominant
error, lighter ions seem to currently have a lower overall error
due to photon scattering. Also, because both errors are a function
of $M\omega_{\text{trap}}$ lighter ions will reach the same error
level and power requirements with a higher $\omega_{\text{trap}}$
when compared to heavier ions. This allows for faster gate
operation. For a given trap geometry and applied potentials, the
axial trap frequency scales as $1/\sqrt{M}$ and the radial trap
frequencies scale as $1/M$. For the ion crystal to remain along
the axial trap direction the radial frequencies have to be larger
than the axial frequencies; therefore the limiting frequency is
the radial frequency. In this case, $M\omega_{\text{trap}}$ is
independent of the ions' mass. Heavier ions under these conditions
will have a lower $\omega_{\text{trap}}$, leading to slower gate
operation.

In conclusion, to minimize the effect of scattering on the
fidelity of trapped-ion-qubit gates, one needs to strike a balance
between the desirable characteristics of long wavelength, light
mass, the availability of high power laser sources and, if
possible, the lack of low lying D levels when choosing a specific
ion as a quantum information carrier.

Finally, we remind the reader that here we have focused on
hyperfine ion-qubits and gates that rely on off-resonant Raman
transitions applied in a continuous pulse. Other kinds of
trapped-ion gates or qubits could have different limitations on
the gate fidelity due to spontaneous photon scattering.

We thank E. Knill for many insightful discussions and comments, J.
Home, J. Amini and A. S{\o}rensen for helpful comments. This work
is supported by DTO and NIST. Contribution of NIST; not subject to
US copyright.

\appendix
\section{Dephasing vs. entanglement with different photon modes}\label{appendix1} In section \ref{section3} we calculated the error due to Rayleigh
photon scattering without considering the scattered photon degrees
of freedom. Rather, we considered dephasing due to the random
phase that is generated by the scattered photon's recoil.

As noted, we can also view the gate error caused by Rayleigh
scattering of photons as arising from the entanglement between the
photon and the ion-qubit external degrees of freedom. For example,
a photon that is scattered during the gate, while the two spin
wave packets are displaced from each other, can be collected by an
ideal imaging system. Since the two-ion collective spin is
entangled with their position, the position at which the photon is
detected at the image plane can give ``which-way'' information,
thereby measuring the ions' spin, collapsing the entanglement, and
causing an error in the gate. Experiments investigating atomic
decoherence due to this effect were performed in neutral-atom
interferometers \cite{Pritchard95, Weitz01}. This effect was also
calculated in \cite{Feagin06} for a single ion in a superposition
of two different motional coherent states.

The general equivalence of these two points of view (dephasing vs.
entangling with the environment) was explained in \cite{Stern90}.
Here we show this equivalence for the trapped ion case where the
ion is undergoing an (ideally) spin-dependent closed loop
displacement. For simplicity we examine the case of a single ion;
it is, however, straightforward to generalize the following proof
to the case of two ions.

We first calculate the final state fidelity by examining
entanglement with the photon modes. As in the two-qubit gate, a
single ion is prepared in an equal superposition of spin states
and is cooled to the motional ground state:
\begin{equation}
|\Psi\rangle = \big(\frac{1}{\sqrt{2}}|\uparrow\rangle +
\frac{1}{\sqrt{2}}|\downarrow\rangle\big)\otimes |0\rangle_M
\underset{\textbf{k}}\otimes |0\rangle_\textbf{k}. \label{Appendix1}
\end{equation}
Here kets with subscript $M$ represent motional states, and kets
with the subscript $\textbf{k}$ the different photon modes, which
are initially empty (neglecting the laser mode). The ion is
subsequently driven by an oscillating force which is detuned from
its motional resonance. The direction of force is opposite for the
two different spin states. Ideally the two spin states would
traverse opposite circular trajectories in phase space (in a frame
rotating at the motional mode frequency). At the end of the gate
drive, both parts of the superposition (ideally) return to the
ground state of motion, and acquire the same geometric phase
$\phi_{\text{gate}}$ which is proportional to the phase-space area
encircled. The state at the end of the ideal gate drive is
\begin{equation}
|\Psi_{\text{ideal}}\rangle =
e^{i\phi_{\text{gate}}}\big(\frac{1}{\sqrt{2}}|\uparrow\rangle +
\frac{1}{\sqrt{2}}|\downarrow\rangle\big)\otimes |0\rangle_M
\underset{\textbf{k}}\otimes |0\rangle_\textbf{k}.
\label{Appendix1a}
\end{equation}

Now assume the ion scattered a photon during the gate drive.
Immediately before the scattering event, the gate evolution
produces
\begin{equation}
|\Psi\rangle =
\big(\frac{1}{\sqrt{2}}e^{i\phi}|\uparrow\rangle|\alpha\rangle
+\frac{1}{\sqrt{2}}e^{i\phi}|\downarrow\rangle|-\alpha\rangle\big)
\underset{\textbf{k}}\otimes|0\rangle_\textbf{k}, \label{Appendix2}
\end{equation}
where $\pm\alpha$ are the conditional phase-space displacement for
the two spin states (we have dropped the subscript $M$) and $\phi$
is the geometric phase accumulated by the time of scattering.
Immediately after scattering, a single photon is created in mode
$\textbf{k'}$ and the ion correspondingly recoils:
\begin{eqnarray}
&& |\Psi\rangle = \\ \nonumber
&&\frac{1}{4\pi}\int\delta(E_\textbf{k'})\big(\frac{1}{\sqrt{2}}e^{i\phi}e^{i\textbf{q'}\cdot
\hat{\textbf{r}}}|\uparrow\rangle|\alpha\rangle +
\frac{1}{\sqrt{2}}e^{i\phi}e^{i\textbf{q'}
\cdot\hat{\textbf{r}}}|\downarrow\rangle|-\alpha\rangle\big)\\
\nonumber
&&\otimes|1\rangle_\textbf{k'}\underset{\textbf{k}\neq\textbf{k'}}\otimes|0\rangle_\textbf{k}d\textbf{k'}.
\label{Appendix3}
\end{eqnarray}
Here $\bf{q'}=\bf{k'}-\bf{k_L}$ is the wave vector difference
between the scattered photon and the laser beam. The
delta-function $\delta(E_\textbf{k'})$ enforces energy
conservation. The electromagnetic field is now represented by a
superposition of states in which a single photon was scattered
into a certain mode while all other modes are empty. Each part of
this superposition is correlated with the corresponding momentum
recoil operator acting on the trapped ion. Recoil into directions
other than the trap axis will give rise to motion which is common
to both parts of the superposition, and can therefore be traced
over. We, therefore, neglect recoil into the dimensions other than
the trap axis direction, and approximate $e^{i\textbf{q'}\cdot
\hat{\textbf{r}}}\simeq e^{iq'_z\hat{z}}$, where
$q'_z=\textbf{q'}\cdot\textbf{z}$. By use of Eq.
(\ref{recoil_operator2}) this momentum recoil can be written as a
phase space displacement $e^{i\textbf{q'}\cdot
\hat{z}}=\hat{D}(\beta')$, where $\beta'=iq'_zz_0$, and $z_0$ is
the root mean-square of the ground state spatial spread. Using Eq.
(\ref{displacements commutation}) we add the recoil to the gate
displacement and the state after scattering can be written as
\begin{eqnarray}
&& |\Psi\rangle = \\ \nonumber
&&\frac{1}{4\pi}\int\delta(E_\textbf{k'})\big(\frac{1}{\sqrt{2}}e^{i\phi}e^{i\text{Im}(\beta'\alpha^*)}|\uparrow\rangle|\alpha+\beta'\rangle
\\ \nonumber
&&+\frac{1}{\sqrt{2}}e^{i\phi}e^{-i\text{Im}(\beta'\alpha^*)}|\downarrow\rangle|-\alpha+\beta'\rangle\big)\otimes|\textbf{k'}\rangle
d\textbf{k'}. \label{Appendix4}
\end{eqnarray}

Since the total gate displacement is equal to zero, the remaining
part of the gate displacement is $\mp\alpha$ (depending of the
spin state). Using Eq. (\ref{displacements commutation}) again, we
can write the state of the system after the gate has completed
\begin{eqnarray}
&& |\Psi\rangle = \\ \nonumber
&&\frac{1}{4\pi}\int\delta(E_\textbf{k'})\big(\frac{1}{\sqrt{2}}e^{i\phi_{\text{gate}}}e^{-i\text{Im}(\alpha\beta'^{\star}-\alpha^{\star}\beta')}|\uparrow\rangle|\beta'\rangle
\\
\nonumber
&&+\frac{1}{\sqrt{2}}e^{i\phi_{\text{gate}}}e^{i\text{Im}(\alpha\beta'^{\star}-\alpha^{\star}\beta')}|\downarrow\rangle|\beta'\rangle\big)\otimes|\textbf{k'}\rangle
d\textbf{k'}. \label{Appendix5}
\end{eqnarray}
We can now define the gate phase-error $\Delta\phi' =
\text{Im}(\alpha\beta'^{\star}-\alpha^{\star}\beta')$, and write
the corresponding density operator\
\begin{widetext}
\begin{eqnarray}
&&\hat{\rho}=\frac{1}{2(4\pi)^2}\int\int\delta(E_\textbf{k'})\delta(E_\textbf{k''})\big(e^{-i\Delta\phi'}e^{i\Delta\phi''}|\uparrow\rangle\langle
\uparrow|+
e^{i\Delta\phi'}e^{-i\Delta\phi''}|\downarrow\rangle\langle
\downarrow|+ \\ \nonumber
&&e^{-i\Delta\phi'}e^{-i\Delta\phi''}|\uparrow\rangle\langle
\downarrow|+
e^{i\Delta\phi'}e^{i\Delta\phi''}|\downarrow\rangle\langle
\uparrow|\big)\otimes|\beta'\rangle\langle\beta''|\otimes|\textbf{k'}\rangle\langle\textbf{k''}|d\textbf{k'}d\textbf{k''}.
 \label{Appendix6}
\end{eqnarray}
\end{widetext}
Since we have no information about the mode into which the photon
was scattered, we reduce the above density matrix with a trace
over the photon modes, using $\text{tr}(|\textbf{k'}\rangle\langle
\textbf{k''}|)=4\pi\delta(\textbf{k'}-\textbf{k''})$,
\begin{widetext}
\begin{equation}
\hat{\rho}=\frac{1}{8\pi}\int\delta(E_\textbf{k'})(\frac{1}{2}|\uparrow\rangle\langle
\uparrow|+\frac{1}{2}|\downarrow\rangle\langle
\downarrow|+\frac{1}{2}e^{-i2\Delta\phi'}|\uparrow\rangle\langle
\downarrow|+\frac{1}{2}e^{i2\Delta\phi'}|\downarrow\rangle\langle
\uparrow|)\otimes|\beta'\rangle\langle\beta'|d\textbf{k'}.\label{Appendix7}
\end{equation}
\end{widetext}
Note that the coherences of the density matrix in Eq.
(\ref{Appendix7}) are given by an average of $e^{i\Delta\phi'}$. A
large spread in $\Delta\phi'$ will ``wash-out'' coherence and
leave an incoherent statistical mixture. Tracing over the motional
degrees of freedom we are now ready to evaluate the fidelity with
respect to the ideal gate output state (Eq. (\ref{Appendix1a})).

\begin{equation}
F=\frac{1}{2}+\frac{1}{2}\langle\cos(2\Delta\phi')\rangle=\langle\cos^2\Delta\phi'\rangle,
\label{Appendix8}
\end{equation}
where averaging is performed with respect to all of the photon
modes:
\begin{equation}
\langle\cos^2\Delta\phi'\rangle =
\frac{1}{4\pi}\int\delta(E_\textbf{k'})\cos^2\Delta\phi'
d\textbf{k'}. \label{Appendix9}
\end{equation}

We now turn to calculate the fidelity looking only at dephasing as
in Sec. \ref{recoil_displacement_error}, i.e. loss of fidelity due
to the photon scattering random phase. The gate output state is
then
\begin{equation}
|\Psi_{\text{ideal}}\rangle=e^{i\phi_{\text{gate}}}\left(\frac{1}{\sqrt{2}}|\uparrow\rangle
+ \frac{e^{i\Delta\phi}}{\sqrt{2}}|\downarrow\rangle\right),
\label{Appendix10}
\end{equation}
where $\Delta\phi$ is the random scattering phase (Eq. \ref{phase
error2}). The fidelity of this state with respect to the ideal
gate output state is identical to that calculated considering
entanglement with the photon modes in Eq. (\ref{Appendix8}).


\begin{references}
\bibitem{gRemark} Here we assume, that $\gamma$ is the same for
the $P_{3/2}$ and the $P_{1/2}$ levels. This is a reasonable
assumption for most ions considered. For the heavier ions in Table
\ref{Table0} the difference can be $10$ $\%$ to $30$ $\%$. The
values for $\gamma$ given in the table and used in the
calculations are those of the $P_{1/2}$ level.
\bibitem{D1 remark} The photon density of states and, hence, the scattering rate will change as the emission wavelength varies with the laser
detuning. For most ions considered the fine-structure splitting is
 small compared to the distance between the $P$ and the $D$
levels and therefore this will be a small effect. This is not the
case however for $^{199}$Hg$^+$ and $^{171}$Yb$^+$, and therefore
for these ions this evaluation of $\epsilon_{D}$ is only
approximate.
\bibitem{D2 remark} In situations where the detuning is not large compared to $\omega_\text{f}$ it would be more accurate to average the branching ratios of the two P levels with weights corresponding to their excitation
probaility. For most ions considered here (excluding $^{199}$Hg$^+$
and $^{171}$Yb$^+$) the difference in branching ratios is small; we,
therefore, take a non-weighted average.
\bibitem{multiple scattering remark} For realistic parameters $P_{\text{scat}}<<1$; therefore we
neglect the possibility that more than one photon is scattered
during the gate.
\bibitem{No Correlation Remark} We assume that the distance between the ions is large compared to the wavelength of light. Therefore, scattering events of the two ions are not correlated.
\bibitem{heavier ions}Since the off-resonant coupling to the carrier scales as $1/\eta$,
even with our choice of constant $\omega_{\text{trap}}$ lighter
ions suffer less from errors due to this coupling. Furthermore,
lighter ions require lower RF trap potentials for the same
$\omega_{\text{trap}}$.
\bibitem{worstcase_vs_average} For a small phase-error, the
error averaged over all possible input states is a factor of 2 lower
than the worst-case input error.
\end{references}
\end{document}